\documentstyle[twocolumn,prb,aps,psfig,epsf,floats]{revtex}
\begin{document}
\title{Nature of the quantum phase transitions in the two-dimensional 
hardcore boson model}
\author{F. H\'ebert, G. G. Batrouni}
\address{
Institut Non-Lin\'eaire de Nice, Universit\'e de Nice--Sophia
Antipolis, 1361 route des Lucioles, 06560 Valbonne, France}
\author{R. T. Scalettar}
\address{Physics Department,
University of California,
Davis CA 95616, USA}
\author{G. Schmid, M. Troyer}
\address{Theoretische Physik, Eidgen\"ossische Technische Hochschule
Z\"urich, CH-8093 Z\"urich, Switzerland}
\author{A. Dorneich}
\address{Institut f\"ur Theoretische Physik, Universit\"at W\"urzburg,
97074 W\"urzburg, Germany}
\address{\mbox{}}
\address{\parbox{14cm}{\rm \mbox{ } \mbox{ }
We use two Quantum Monte Carlo algorithms to map out the phase diagram
of the two-dimensional hardcore boson Hubbard model with near ($V_1$)
and next near ($V_2$) neighbor repulsion. At half filling we find
three phases: Superfluid (SF), checkerboard solid and striped solid
depending on the relative values of $V_1$, $V_2$ and the kinetic
energy. Doping away from half filling, the checkerboard solid
undergoes phase separation: The superfluid and solid phases co-exist
but not as a single thermodynamic phase. As a function of doping, the
transition from the checkerboard solid is therefore first order.  In
contrast, doping the striped solid away from half filling instead
produces a striped supersolid phase: Co-existence of density order
with superfluidity as a single phase.  One surprising result is that
the entire line of transitions between the SF and checkerboard solid
phases at half filling appears to exhibit dynamical $O(3)$ symmetry
restoration. The transitions appear to be in the same universality
class as the special Heisenberg point even though this symmetry is
explicitly broken by the $V_2$ interaction.}}
\address{\mbox{}}
\address{\parbox{14cm}{\rm \mbox{ } \mbox{ }
}}
\maketitle

\section{Introduction}

The boson Hubbard Hamiltonian\cite{FisherM} has been studied as a
model of the superconductor--insulator transition in materials with
preformed Cooper
pairs,\cite{cha,sorensen,ourprb5,runge,elesin,kampfzim} of Helium in
disordered and restricted geometries,\cite{ourprl2,ourprb4,kanwal} of
spin--flop transitions in quantum spin systems in external magnetic
fields,\cite{spinflop} and of supersolid behavior.\cite{ggb3,ggb4} As
with the fermion Hubbard Hamiltonian, the boson model explores the
role of correlations in inducing ordered phases of many quantum
mechanical particles, and the nature of the quantum phase transitions
between these phases.  However, unlike the fermion case where it is
very difficult to reach low temperatures away from points of special
particle--hole symmetry, Quantum Monte Carlo (QMC) simulations of the
doped boson system have no ``sign problem'' and hence can successfully
be performed.  Many fascinating and unexpected features arise, for
example re-entrant Mott insulating behavior\cite{monien}, universal
conductivity,\cite{FisherM,cha,sorensen,ourprb5,runge,cha2} and
supersolidity.\cite{ggb3,ggb4} Indeed, until algorithms are developed
to deal with the sign problem in fermion QMC the boson Hubbard
Hamiltonian offers us the best opportunity to explore systematically
the details of the competition between phases with diagonal and
off--diagonal long range order with QMC simulations.

In this paper we extend some of the previous work which established
the basic phase diagram of the
model\cite{FisherM,ourprl1,ourprb1+2+3,Trivedi1} in order to
characterize the detailed critical properties of the transitions
between the phases.  We will use two recently developed algorithms.  The
first\cite{ggb1,ggb2} is based on a duality transformation which
enables an exact mapping of the boson Hubbard Hamiltonian onto a model
of conserved currents, and the second\cite{sandvik,troyer} is based on
a stochastic series expansion of the imaginary time evolution
operator.

The paper is organized as follows: We will first review the definition
of the boson Hubbard model and some of its basic qualitative
properties.  We will then provide a brief review of the two numerical
algorithms we employ.  Our presentation of the results begins with a
discussion of the half--filled system which focusses on a scaling
analysis of the transition from a strong coupling solid to a weak
coupling superfluid phase.  Next, away from half--filling, we discuss
the nature of the coexistence of the solid and the superfluid.
Finally, we present some concluding remarks and open questions.

\section{The Boson Hubbard model}

The hardcore boson Hubbard Hamiltonian is,
\begin{eqnarray}
H=&-&t\sum_{\langle
   {\bf i},{\bf j}\rangle}(a_{{\bf i}}^{\dagger}a_{{\bf j}}+
a_{{\bf j}}^{\dagger}a_{{\bf i}})
\nonumber \\
&+&V_{1}\sum_{\langle {\bf i},{\bf j} \rangle}
   n_{{\bf i}} n_{{\bf j}}+
V_{2}\sum_{\langle\langle {\bf i},{\bf k}\rangle\rangle}
   n_{{\bf i}} n_{{\bf k}}.
\label{hub-ham}
\end{eqnarray}
$a_{\bf i}$ ($a_{\bf i}^{\dagger}$) are destruction (creation)
operators of hard--core bosons on site ${\bf i}$ of a 2--d $L\times L$
square lattice, and $n_{\bf i}$ is the density at site ${\bf i}$.  The
hopping parameter is chosen to be $t=1$ to fix the energy scale.
$V_1$ ($V_2$) is the near (next near) neighbor interaction.

For weak couplings, the ground state of the Hamiltonian is a
superfluid.  Increasing the near neighbor interaction strength, $V_1$,
at half filling drives a transition to a checkerboard solid phase
where the sites are alternately occupied and empty.  This phase is
characterized by a vanishing superfluid density, $\rho_s$, long range
density--density correlations, and a gap in the energy spectrum
reflected, for example, as a vanishing compressibility, $\kappa$, and
corresponding plateau in a plot of density $\rho$ versus chemical
potential $\mu$.  Increasing the next near neighbor interaction
strength $V_2$ at half filling can likewise drive a transition to a
striped solid, where horizontal (or vertical) lines of sites are
alternately occupied.  This phase also has $\rho_s=0, \kappa=0$ and
long range density--density correlations.

At $V_2=0$, and after an appropriate sublattice spin rotation, the
hard--core boson model is equivalent to the spin--$\frac12$
antiferromagnetic XXZ model, with half--filling corresponding to the
zero magnetization sector.  The hopping $t$ maps onto $J_x/2$ while
the interaction strength $V_1$ maps onto $J_z$ and the chemical
potential $\mu$ is related to the magnetic field as $h = \mu-zV_1/2$,
where $z$ is the number of nearest neighbors of a lattice site.  In
this language, superfluid order corresponds to magnetic order in the
XY plane, while density order corresponds to magnetic order in the Z
direction.  The boson superfluid--cdw insulator phase transition at
$V_1=2t$ corresponds to the XY--Ising change in universality class at
$J_x=J_z$.  At this, the Heisenberg point, the Hamiltonian has an
$O(3)$ symmetry and as a consequence the critical temperature is
driven to $T_c=0$.  This symmetry is explicitly broken for other
values of $V_1$ or for nonzero $V_2$. The boson--Hubbard model with
nonzero $V_2$ also has a spin analog, namely to a Hamiltonian with
next--near neighbor exchange.

The behavior of the boson--Hubbard model away from half-filling is
considerably more complex.  While the compressibility surely becomes
nonzero, so that the state is not a Mott insulator, it is possible
that, despite doping, the charge correlations remain long ranged.  If
the doped holes are mobile and interspersed with the density ordered
bosons, one has simultaneous superfluid order.  On the other hand, the
doped holes might phase separate leaving distinct regions of the
lattice with superfluid and charge ordering.  We will carefully
explore these alternate possibilities and describe the nature of the
transitions between them.

\section{The Monte Carlo Algorithms}

Until relatively recently, algorithms to simulate interacting quantum
bosons on a lattice suffered significant weaknesses.  Most
importantly, they had problems with extremely long autocorrelation
times, which, as in classical Monte Carlo, were caused by the
inability of local changes to move configurations effectively through
phase space.  A related difficulty was that local moves also resulted
in global conservation laws which limited the accessible regions of
phase space and hence the measurements that could be performed.  In
the last several years, ``cluster'' and ``loop'' algorithms have very
successfully addressed some of these problems.\cite{loop}
Unfortunately, these algorithms do not work equally well in all
regions of parameter space, notably away from half filling. In
addition, they are not simple to implement when the Hamiltonian is
made more complex, e.g.~with the inclusion of disorder or longer range
interactions.\cite{dishubsims}
In this section we review two approaches which have
short autocorrelation times but also are easily implemented for these
more general models.

\underbar{Dual Algorithm:}
To perform efficient boson simulations we use a newly developed QMC
algorithm based on the exact duality transformation of the Bosonic
Hubbard model.\cite{ggb1,ggb2} This approach begins by expressing the
partition function as a path integral over coherent states.  To
implement the duality transformation exactly we followed the method of
Ref.~\onlinecite{ggbdual} The result is that hardcore bosons are
represented by conserved ``currents'' that propagate in the positive
imaginary time direction and which can make jumps in any spatial
direction.  The path integral is transformed into a sum over all
deformations of the currents, very much like the world line
algorithm. This formulation is similar to that of
Refs.~\onlinecite{cha,sorensen} in that it relies on the duality
transformation, but different in that this transformation is exact in
our case. Consequently, with our algorithm we simulate the true
hard-core Hubbard model, whereas Refs.~\onlinecite{cha,sorensen} study
the very high density (many bosons/site) limit.

The simulation is done in the standard way. A deformation is suggested
and it is accepted or rejected in a way satisfying detailed balance. A
feature of this algorithm which is not easily implemented in the world
line algorithm is that the imaginary time step is not subdivided
by a checkerboard break--up, and
so a particle can hop several lattice sites at a time. This has the
great advantage that it enables us to measure the correlation function
of the superfluid order parameter, $\langle a^{\dagger}_{\bf i} a_{\bf
j}\rangle $, for large $|{\bf i}-{\bf j}|$, which is very difficult to
measure efficiently in older approaches.

\underbar{Stochastic Series Expansion:} The stochastic series
expansion (SSE) algorithm with operator loop
update\cite{sandvik,troyer} was used for additional calculations on
larger systems in the grand canonical ensemble. This algorithm does
not suffer from time discretization errors and is one of the most
effective algorithm for quantum systems. It has been shown to be as
effective as the loop algorithm \cite{loop} for models where the loop
algorithm can be applied. In bosonic models it is better than the loop
algorithm, since it does not suffer from the exponential slowing down
of the loop algorithm when the chemical potential is tuned away from
half filling.\cite{troyer}.

\underbar{Measurements:}
The energy $E$ is obtained as the expectation value of the
Hamiltonian and is used to determine the
chemical potential via the relation
$\mu=\partial E/\partial N$, from simulations
in the canonical ensemble.  Hysteretic behavior
in $E$ also provides supporting evidence in characterizing the
order of phase transitions.

It is straightforward to measure the density-density correlations
and their Fourier transform, the structure factor:
\begin{eqnarray}
c({\bf l})&=&\langle n_{{\bf j}+{\bf l}} n_{\bf j} \rangle
\nonumber \\
S({\bf q}) &=&  \sum_{\bf l} e^{i {\bf q} \cdot {\bf l}} c({\bf l}).
\label{corrfct}
\end{eqnarray}
These quantities characterize the diagonal long range order.  Divergence 
of
$S(\pi,\pi)$ in the thermodynamic limit
indicates checkerboard order, and in $S(\pi,0)$ or
$S(0,\pi)$ striped order.  On finite lattices, the structure factor at
the appropriate momentum diverges as the system size in the ordered
phase, so that a scaling analysis of simulations on different lattice
sizes can demonstrate long range order.  From the density itself comes
the compressibility which characterizes Mott insulating behavior.

It is also crucial to obtain the winding number, since it will be used
to determine the superfluid density $\rho_s \propto \langle W^2
\rangle$, a relationship first emphasized in the context of Quantum
Monte Carlo simulations by Ceperley and Pollock.\cite{CP} While global
conservation laws on winding and particle numbers preclude the
straightforward evaluation of the superfluid density in the dual
algorithm and other traditional world--line approaches, we used
methods which circumvent this difficulty.\cite{ourprl1,ggb1}
Specifically, we calculate the ``pseudo'' current-current correlation
function at different imaginary times. The ``pseudocurrent'' is
defined as the net number of bosons which jumped in a given direction
in one imaginary time step. It can be shown easily that the Fourier
transform of this correlation function approaches $\langle W^2
\rangle$ as the (Matsubara) frequency approaches zero.\cite{ourprl1,ggb1}

Meanwhile, the global loop updates of the SSE algorithm allow
modifications of the particle number and of the winding numbers. The
superfluid density $\rho_s$ can thus be measured directly from the
winding number fluctuations in these approaches.  Two of us have
recently developed a method to measure Green's functions like $\langle
a^{\dagger}_{\bf i} a_{\bf j}\rangle$ in the SSE algorithm. We refer
to Ref. \onlinecite{troyer} for details of this method.  Values for
the superfluid density obtained via the dual and SSE algorithms,
although obtained very differently, are entirely consistent.

\section{Phase Diagram and Transitions Half Filling}

As we have already described, at half filling, one can easily
recognize the existence of at least three phases. For weak $V_1$ and
$V_2$ it is clear that a superfluid phase exists and, as we will see
below, this extends to very strong repulsions when the competing
interactions nearly balance. If $V_1$ dominates, the energy cost of
having near neighbors becomes too high and the bosons organize
themselves into a checkerboard solid. On the other hand, when $V_2$
dominates, it is less costly to have near neighbors compared to next
near neighbors and the bosons organize themselves in a striped
solid. What is not obvious is whether there are other phases, for
example supersolids separating the solid phases from the superfluid
phase. In previous work\cite{ggb3,ggb1,ggb2} we demonstrated that at
half filling there are no supersolid phases in this model. This is
confirmed in the present work where we determine the phase diagram
more accurately than before and study in detail the nature of the
transitions.  The case $V_2=0$ and $V_1=2$ serves as a good test for
our simulations, since it corresponds to the well--understood
Heisenberg point of the XXZ model.

The checkerboard and striped solids have structure factors which
diverge as $L^2$, with momentum ordering vectors $(\pi,\pi)$ and
$(\pi,0)$ (or $(0,\pi)$) respectively.  Figure~1 shows these two
quantities for fixed $V_1$ as $V_2$ is varied. We see that as $V_2$ is
increased, $S(\pi,\pi)$ falls to zero, $\rho_s$ takes a finite value
while $S(\pi,0)$ remains zero. This indicates a $(\pi,\pi)$-solid to
superfluid phase transition. Increasing $V_2$ further, $\rho_s$
vanishes abruptly while $S(\pi,0)$ takes a finite value indicating a
superfluid to striped $(\pi,0)$-solid.

\begin{figure}
\psfig{file=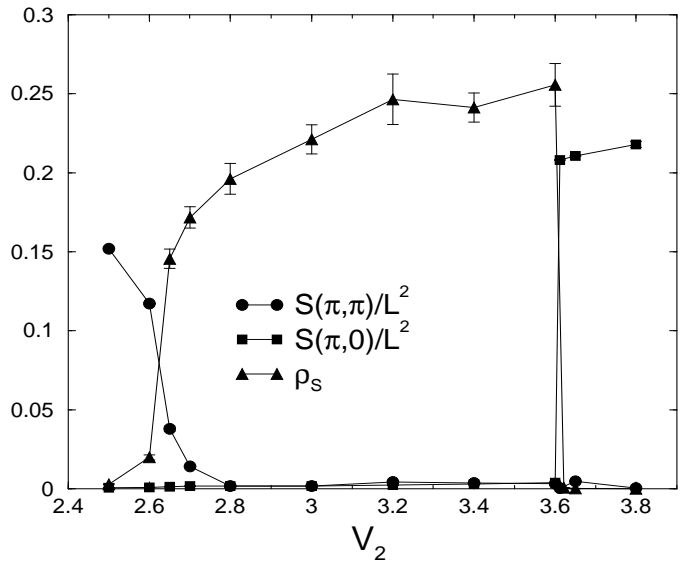,height=3.0in,width=3.5in}
\vskip-00mm
\caption{
The structure factor, $S(\pi,\pi)$ (circles), $S(\pi,0)$ (squares) and
the superfluid density, $\rho_s$ (triangles) at half filling for
$V_1=6$ as functions of $V_2$. $L=12, \beta=12$}
\end{figure}

Putting together the transition points obtained from several such
slices, we arrive at the ground state phase diagram in the $(V_1,V_2)$
plane at half filling. This is shown in Fig.~2.  Two remarks are in
order. It is interesting to note in Fig.~2 that even for very large
values of $V_1$ and $V_2$, there is no direct transition between the
checkerboard and striped phases: The superfluid phase seems always to
intervene. Presumably, the two solid phases eventually meet as both
$V_1$ and $V_2$ go to infinity.  This is in contrast with the mean
field result\cite{ggb3} that the two phases meet at
$(V_1,V_2)=(4,2)$. The second remark is that $V_2$ does not need to be
larger than $V_1$ in order for striped order to win over checkerboard
order. For example, $(V_1,V_2)=(6,4)$ is in the striped phase. This is
important because it makes the striped phase more likely to appear
physically since one might expect near neighbor repulsion to be
stronger than next near neighbor.

\begin{figure}
\psfig{file=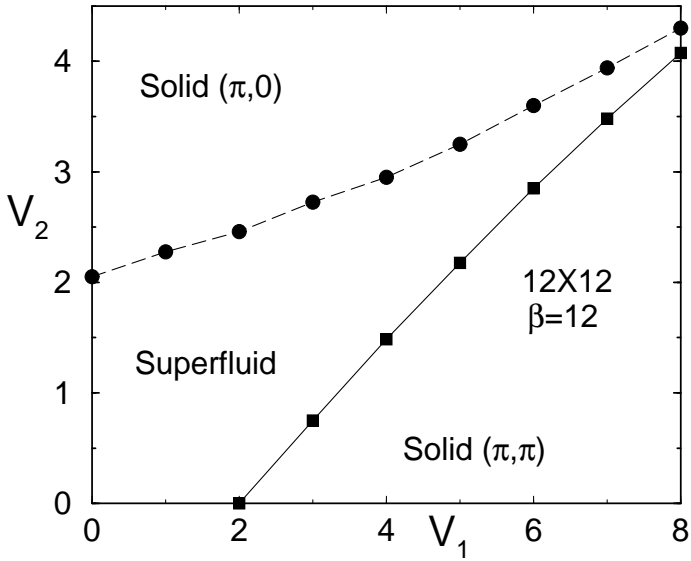,height=3.0in,width=3.5in}
\vskip-00mm
\caption{ The ground state phase diagram in the $(V_1,V_2)$ plane. The
dashed line indicates first order transitions, the solid line exhibits
what appears to be dynamically restored $O(3)$ symmetry except at
$(V_1=2,V_2=0)$ where the symmetry is explicit in the hamiltonian (see
text below). }
\end{figure}

Fig.~1 also shows that $(\pi,\pi)$-SF transition appears smooth
suggesting a continuous phase transition. This will be examined in
detail below. On the other hand, Fig.~1 also shows that the
SF-$(\pi,0)$ transition is very sudden, suggesting a first order
transtion.  To verify the first order nature of the SF-$(\pi,0)$
transition we look for hysteresis in $S(\pi,0)$ and $\langle E
\rangle$ as $V_2$ is increased and then decreased, always starting a
new simulation from the last configuration of the previous run.  The
results are shown in Fig.~3. Hysteresis is clearly seen, supporting
the evidence for a first order transition.  Similar parameter sweeps
through the $(\pi,\pi)$-SF transition exhibit no hysteresis.

\begin{figure}
\psfig{file=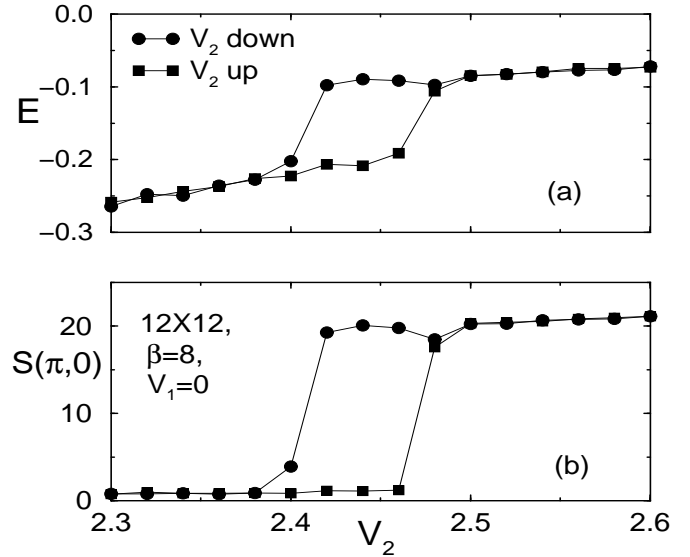,height=3.0in,width=3.5in}
\vskip-00mm
\caption{Hysteresis of (a) the energy $E$ and (b) $S(\pi,0)$,
as $V_2$ is increased and decreased showing the striped solid to
superfluid transition to be first order.}
\end{figure}

To determine the order of the transition from checkerboard solid to
superfluid we do finite size scaling. Since the Heisenberg point
($V_1=2,V_2=0$) is very special (explicit $O(3)$ in the Hamiltonian),
we first did the analysis away from it by fixing $V_1=3$ and finding
the transition as $V_2$ is changed (see Fig.~1).  Since from Fig.~1,
the SF to ($\pi,\pi$)--solid transition appears continuous, we will
first proceed by assuming a second order transition and carrying out
the appropriate finite size scaling.  What this analysis will show is
that while a quite reasonable data collapse can be achieved, the
requisite dynamic critical exponent is anomalously small.  This will
lead us to perform a more careful analysis which will reveal the true,
and more subtle, critical behavior.

In Fig.~4 we show $\rho_s$ versus $V_2$ for system sizes $L=6,8, 10,
12$.  In a quantum phase transition, the finite size scaling function
not only depends on the appropriately scaled distance to the critical
point, but also on the ratio of the lattice dimensions in space and
imaginary time.  It is standard to assume the following form for the
superfluid density
\cite{FisherM},
\begin{equation}
\rho_s \propto \frac{1}{L^z} F\left( \frac{V_2 -V_2^c}{L^{-1/\nu}},
\beta/L^z \right).
\label{FSS}
\end{equation}
One approach is to simulate several sets of lattice sizes, each set
with a different space/imaginary time aspect ratio associated with
different guesses for $z$.  Instead, we first use another approach
which is to choose the inverse temperature large enough ($\beta = 20$
usually suffices) so that the second argument in $F$ is a constant as
$L$ is varied.  With this choice, $\rho_s L^z$ must then be
independent of $L$ at the critical point.  The best intersection was
found when $z=0.25$, giving a critical interaction value $V_2^c =
0.765$ (see Fig.~4).  Replotting this figure using the scaled variable
$(V_2 -V_2^c) L^{1/v}$ we obtain very good data collapse shown in
figure Fig.~5.  This collapse yields the value of the correlation
function exponent $\nu=0.36$.

Once these critical exponents were found, we redid the simulations but
with the two parameter finite size scaling analysis mentioned above in
which the temperature is varied with $L$, $\beta \propto L^z$, in
order to keep the second argument in Eq.~\ref{FSS} constant. This
reproduced the same values found for $z$ and $\nu$.

\begin{figure}
\psfig{file=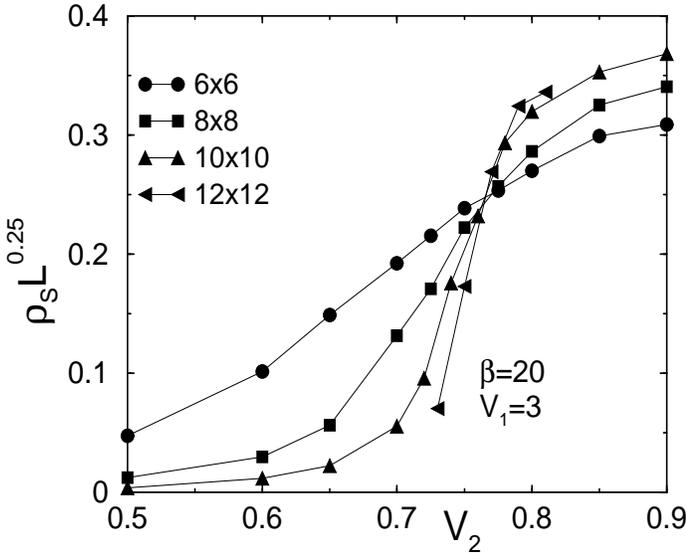,height=3.0in,width=3.5in}
\vskip-00mm
\caption{
$\rho_s$ as a function of $V_2$ for $L=6, 8, 10, 12$. The intersection
gives the critical value $V_2^c=0.765$.  }
\end{figure}

\begin{figure}
\psfig{file=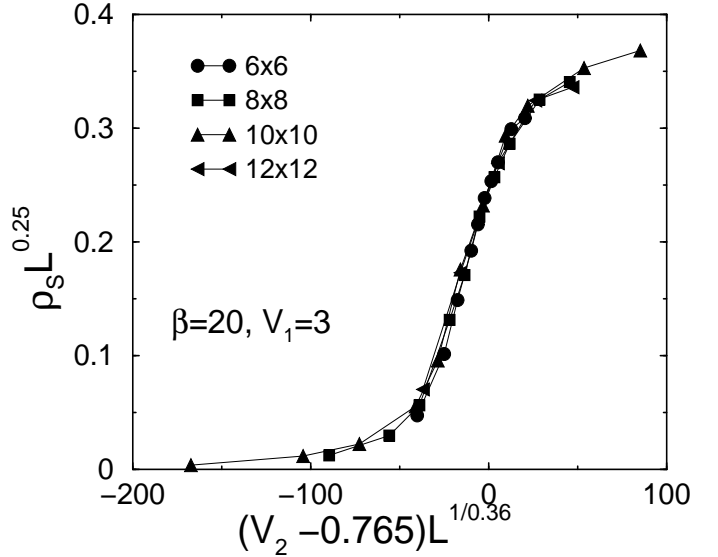,height=3.0in,width=3.5in}
\vskip-00mm
\caption{
Same as Fig.~4 but using the scaling variables $\rho_s L^{0.25}$ and
$|V_2-V_2^c|^{1/\nu}$. This yields the values $z=0.25$ and $\nu=0.36$.
}
\end{figure}

Scaling analyses and data collapse of this level of quality are rather
commonly used to draw conclusions concerning the appropriate critical
behavior and universality. However, the value of the dynamical
critical exponent, $z=0.25$, is surprisingly small. All previous
debate\cite{ggb3,vanott} had been whether $z=1$ or $z=2$.  This leads
us to re--examine the process and, in particular, check the validity
of the above finite size scaling analysis, by applying exactly the
same assumptions and methodology to the transition at the Heisenberg
point, where we know they should not hold.  We obtain very similar
scaling data, with the same values of $z$ and $\nu$.

The similarity between the behavior of the superfluid density near the
Heisenberg point with that at finite $V_2$ suggests that the entire
line of phase transitions separating the SF and ($\pi,\pi$)--solid
phases might be in the same class as the Heisenberg point.  However,
especially given the possibility of generating `acceptable' finite
size scaling plots despite the known critical behavior, that we have
just demonstrated, this conjecture clearly requires careful testing,
which we shall now describe.

The Heisenberg point is very special in that the Hamiltonian has an
$O(3)$ symmetry which is explicitly broken everywhere else in the
$(V_1,V_2)$ plane. In order for the transition line from the SF to the
($\pi,\pi$)--solid to be in the same universality class, this symmetry
must be dynamically restored at the transition. To check this
numerically, we first need to determine accurately a transition point
away from the special Heisenberg point. To this end, we fixed $V_1=3$
and studied, for many values of $V_2$ and $L$, the behavior of
$S(\pi,\pi)$ and the condensate, $N({\vec k=0}) = {\tilde
a}^{\dagger}({\vec k=0}) {\tilde a}({\vec k=0})$ where ${\tilde a}$ is
the Fourier transform of the destruction operator. The results for
$N({\vec k}=0)$ are shown in Fig.~6 and indicate that the transition
happens at $0.760 < V_2 < 0.762$ for $V_1=3$. Furthermore, a
substantial jump in $N(k)$ is indicated over this narrow window of
parameters.  This very abrupt transition argues for a discontinuous
transition like that at the Heisenberg point.

A sensitive test of this suggestion is that if the $O(3)$ symmetry is
indeed restored at the critical $V_2$, the transition can only take
place at zero temperature. To verify this we did simulations at finite
temperature to determine the transition temperatures as a function of
$V_2$. For $V_2< 0.761$, the transition is from a normal bosonic
liquid to the $(\pi,\pi)$ solid, and is expected to be in the 2d Ising
universality class. The transition temperatures were determined from
crossing points of the 4-th order Binder cumulant ratios $1-\langle
S(\pi,\pi)^2\rangle/3\langle S(\pi,\pi)\rangle^2$ for different system
sizes $L$, and plotted in Figure~7.  For $V_2> 0.761$, the transition
is from a normal bosonic liquid to a superfluid, and is expected to be
in the 2d XY universality class. The critical temperaturs $T_{KT}$ of
this phase were determined from the universal jump of the superfluid
density $\rho_s$ at the critical temperature.  In Fig.~7 we see very
clearly that the transition temperature plunges to zero as we approach
the critical $V_2$. In addition, the way the critical temperature
drops to zero is well fitted by $-1/{\rm ln}(V_2-V_{2c})$, just as is
the case at the Heisenberg point.

\begin{figure}
\psfig{file=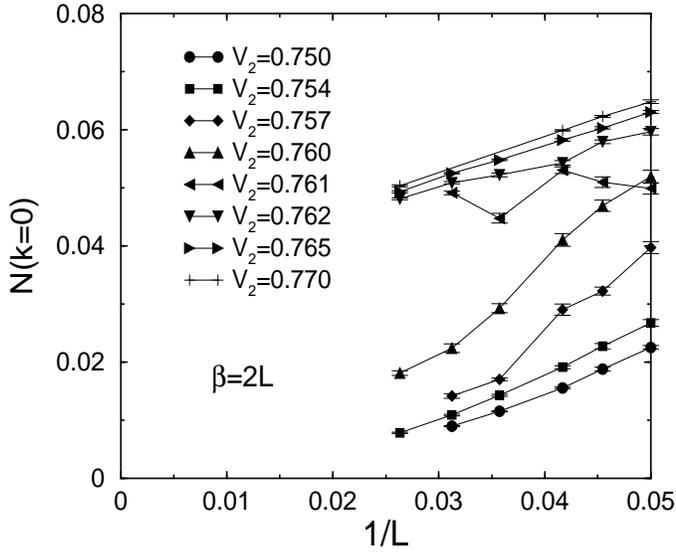,height=3.0in,width=3.5in}
\vskip-00mm
\caption{ The number of bosons in the zero momentum mode, the
condensate, as a function of $1/L$ for different next near neighbor,
$V_2$, repulsion values. $\beta=2L$.
   }
\end{figure}

\begin{figure}
\psfig{file=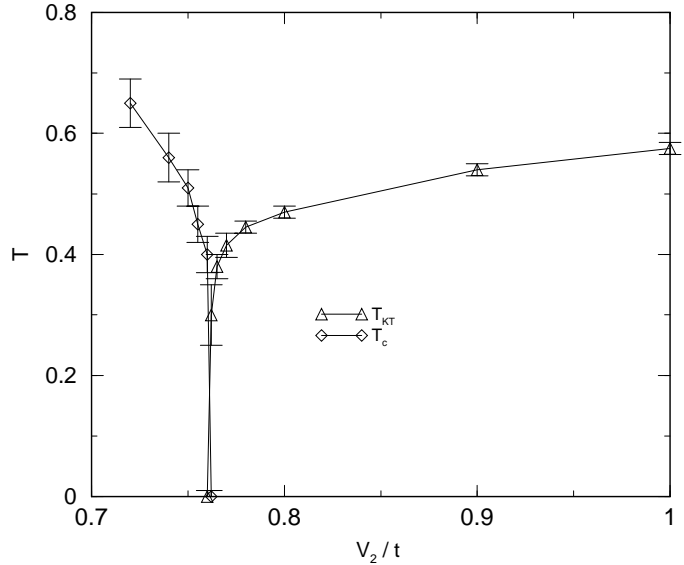,height=3.0in,width=3.5in,angle=-90}
\vskip-00mm
\caption{ The transition temperature, $T$, versus $V_2$. At the
critical $V_2 (\approx 0.761)$ the transition temperature drops to
zero. This suggests possible restoration of the $O(3)$ symmetry.  }
\end{figure}

It seems very likely, therefore, that in the ground state, $T=0$, the
transition at half filling from the checkerboard solid to the
superfluid proceeds via a dynamical restoration of symmetry and is in
the same class as the special Heisenberg point at $V_1=2,
V_2=0$. However, our simulations were done at {\it finite}, albeit
very low, temperatures. Consequently, for the moment, we cannot
exclude the possibility that at very low temperatures the transition
is first order. These issues and the finite temperature phase
transition are currently being studied further.\cite{finiteT}

It is interesting to compare this with what happens in other systems.
In the 2--d extended fermion Hubbard model, there is a similar
competition between an antiferromagnetic ordered phase and a charge
density wave phase.  In the case of fermions, the entire zero
temperature phase diagram in the $(U,V_1)$ (onsite and near neighbor
repulsion) plane at half--filling is believed to be insulating.  There
is no metallic or superconducting region.  This is thought to arise in
part because of the peculiar nature of the two--dimensional Fermi
surface, which has perfect nesting and a logarithmic divergence of the
density of states at half--filling.  Both these features act to
enhance the tendency for diagonal long range order.  However, little
has been pinned down precisely as to the nature of the phase
transitions in two dimensions.  In one dimension, the question has
been addressed, and the transition between the two insulating states
has been argued to be second order at weak coupling and first order at
strong coupling, with an intervening tricritical
point.\cite{HIRSCH,CANNON} Recent work has called this into question,
and suggested that an intervening bond--ordered--wave phase is
important.\cite{BOW}

In the 1--d soft core boson--Hubbard model, the ground state phase
diagram in the $U$ (on--site repulsion) and $V_1$ (near neighbor
repulsion) was also studied extensively\cite{ourprb1+2+3} at filling
$n=1$.  As in the case of 2--d reported here, there is a
weak--coupling superfluid phase which is supplanted by ordered,
insulating Mott and charge density wave phases at large $U$ and $V_1$
respectively.  A study of hysteresis loops and the free energy
barriers indicated that the transition out of the cdw phase to the
superfluid is second order at weak coupling, and that the transition
from cdw to Mott phase at strong coupling is first order.  We see that
there are similarities with the 2--d hardcore case reported here, but
there also differences, namely here there could be a dynamically
restored $O(3)$ symmetry which was explicitly broken in the
hamiltonian.

\section{Doped System}

Now we examine the phase transitions and various phases when the
system is doped away from half filling. The hard--core boson--Hubbard
Hamiltonian, Eq.~1 has particle-hole symmetry. That is, the
transformation $a_{{\bf i}} \rightarrow a_{{\bf i}}^{\dagger}$ maps
$n_{\bf i}
\rightarrow 1-n_{\bf i}$, interchanging occupied and empty sites, but
leaves the Hamiltonian unchanged apart from trivial constants.
Therefore it is sufficient to do the simulations below half-filling
and use this symmetry to calculate physical quantities above.

\subsection{Evolution of Checkerboard Solid}

To explore what happens to the checkerboard solid when the system is
doped, we performed a series of simulations at various values of $V_1$
as the number of bosons in the system is lowered from half filling. We
did this mostly with the canonical dual QMC algorithm where the number
of bosons is fixed and the chemical potential is calculated from the
energy to add a particle to the system,
\begin{equation}
\mu = E(N_{\rm boson}+1)-E(N_{\rm boson})
\label{defmu}
\end{equation}

In Fig.~8 we show a typical result for $V_1=3$. At low densities (in
this case when $\rho<0.4$) we see that $\rho_s$ is finite while
$S(\pi,\pi)$ is small and decreasing as we move away from
$\rho=0.5$. In addition, finite size studies show that $\rho_s$ is
essentially unchanging while $S(\pi,\pi)\to 0$ for a fixed $\rho<0.4$
as $L$ grows. Therefore, this corresponds to a superfluid phase. The
structure factor reaches its maximum at $\rho=0.5$ while $\rho_s$ is
zero there. This is the checkerboard solid discussed in the previous
section.

\begin{figure}
\psfig{file=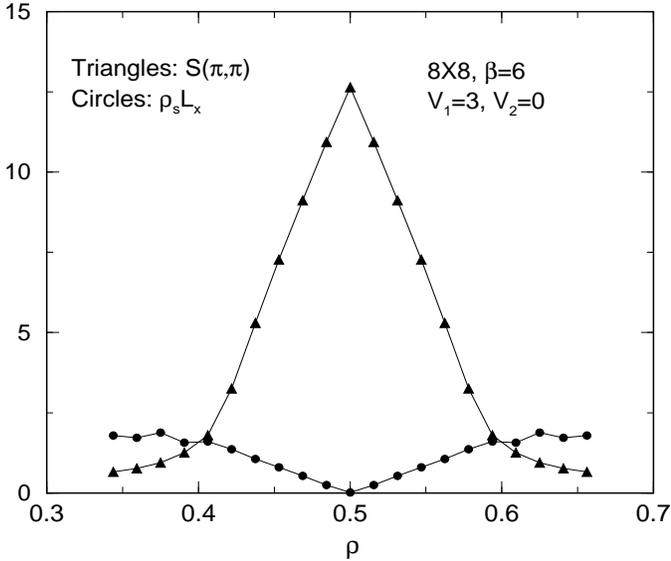,height=3.0in,width=3.5in,angle=-90}
\vskip-00mm
\caption{
The structure factor, $S(\pi,\pi)$, and superfluid density, $\rho_s$,
versus the particle density, $\rho$. For $0.4<\rho<0.5$, both $\rho_s$
and $S(\pi,\pi)$ are finite.
  }
\end{figure}

Between the superfluid phase and the half filling checkerboard solid,
{\it i.e.} for $0.4<\rho<0.5$, Fig.~8 shows that both the structure
factor and superfluid density are non-vanishing. Furthermore, this is
not a finite size effect: For larger systems, $\rho_s$ maintains its
value while $S(\pi,\pi)$ diverges with $L^2$ as it should in the case
of long range density wave order. This, therefore, is a candidate for
a checkerboard supersolid phase.

To verify this possibility, and check the thermodynamic stability of
the supersolid phase we show in Fig.~9 the density, $\rho$, as a
function of the calculated chemical potential, $\mu$. We see that for
all the density values where Fig.~8 shows a supersolid, {\it i.e.}
$0.4<\rho<0.5$, the curve in Fig.~9 has negative slope and therefore
negative compressibility, $\kappa=\partial \rho/\partial
\mu$. Consequently, the apparent checkerboard supersolid phase is not
stable thermodynamically and undergoes phase separation into a mixture
of checkerboard solid and superfluid. This same behavior had
previously been established for the magnetization process of the
spin-${1/2}$ XXZ model on smaller lattices.\cite{spinflop}

\begin{figure}
\psfig{file=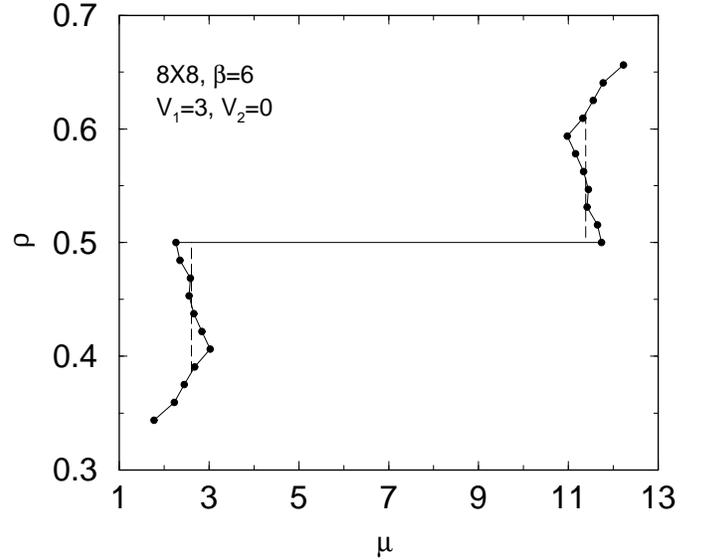,height=3.0in,width=3.5in,angle=-90}
\vskip-00mm
\caption{
The particle density, $\rho$, as a function of the calculated chemical
potential, $\mu$. The slope is the compressibility, $\kappa=\partial
\rho/\partial \mu$.
  }
\end{figure}

To establish this phase separation further, we simulated the system in
the grand canonical ensemble where $\mu$ is the input parameter and
$\rho$ is calculated. If the system undergoes phase separation, as
shown in Fig.~9, then, for the corresponding value of $\mu$, a
histogram of the density should show two peaks, one at $\rho=0.5$ and
the other at $\rho < 0.5$. This is indeed what happens as shown in
Fig.~10 for an $8\times 8$ system at $V_1=2.86$. The simulation is done 
for
several values of the chemical potential. The phase transition takes
place for the $\mu$ value with equal peaks. We verified that the peak
separation does not change when the system size is increased.

\begin{figure}
\psfig{file=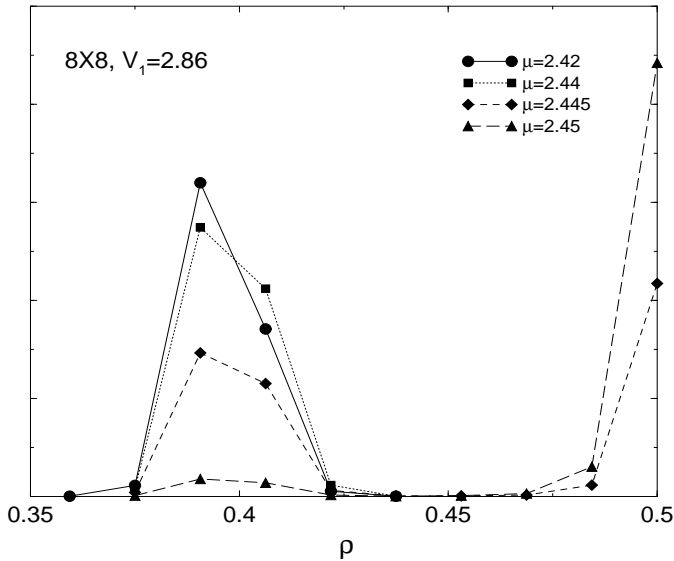,height=3.0in,width=3.5in,angle=-90}
\vskip-00mm
\caption{
Histogram of the particle density as the chemical potential, $\mu$ is
changed. The double peaks show phase separation.
  }
\end{figure}

By repeating the simulations that led to Fig.~9 for various values of
$V_1$, we map out the phase diagram in the ($t/V_1,\mu/V_1$) plane for
$V_2=0$. The transitions between the superfluid and the
($\pi,\pi$)-solid phases are first order except at half filling which
is the special Heisenberg point. This is shown in Fig.~11. This phase
diagram is in agreement with the mean field/spin wave analysis in
Ref.\onlinecite{pich-frey}.

\begin{figure}
\psfig{file=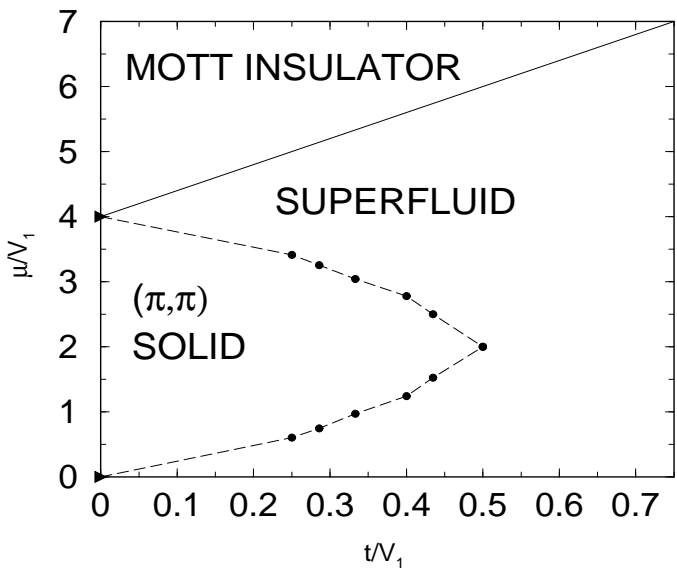,height=3.0in,width=3.5in,angle=-90}
\vskip-00mm
\caption{
The phase diagram for $V_2=0$. The solid line shows the continuous
transition to the Mott phase at full filling, the dashed line shows
the discontinuous first order transitions from the superfluid to the
checkerboard solid at half filling.  The tip of the lobe, $\rho=0.5$,
is the Heisenberg point.  }
\end{figure}

\subsection{Evolution of Striped Solid}

Now we investigate the effect of doping on the striped solid phase
present at half filling.

The top part of Fig.~12 shows the structure factors, $S(\pi,0)$ and
$S(0,\pi)$, for $V_1=0, V_2=5$. For $\rho < 0.3$ we see that the
system is isotropic, $S(\pi,0)=S(0,\pi)$ and vanishing. For $\rho >
0.3$ the symmetry is broken and one of the two vanishes while the
other is large (diverges with the system size). That signals the
formation of stripes along the $x$ or $y$ directions. It is remarkable
that the stripes start forming at such small densities. The figure
also shows that at the same densities where the stripes form, the
superfluid density is no longer isotropic, $\rho_s^x\ne \rho_s^y$. The
superfluid densities in the $x$ and $y$ directions are defined by:
$\rho_s^x=\langle W^2_x\rangle/4t\beta$ and $\rho_s^y=\langle
W^2_y\rangle/4t\beta$, where $W_x$ ($W_y$) is the winding number in
the $x$ ($y$) direction. In addition, the figure shows clearly that
the superlfuid density along the stripes is larger than transverse to
the stripes. Nonetheless, the superfluid density in the transverse
direction does not vanish. Therefore, once again we apparently have a
phase which is both superfluid and solid, thus another candidate for
the supersolid phase.

\begin{figure}
\psfig{file=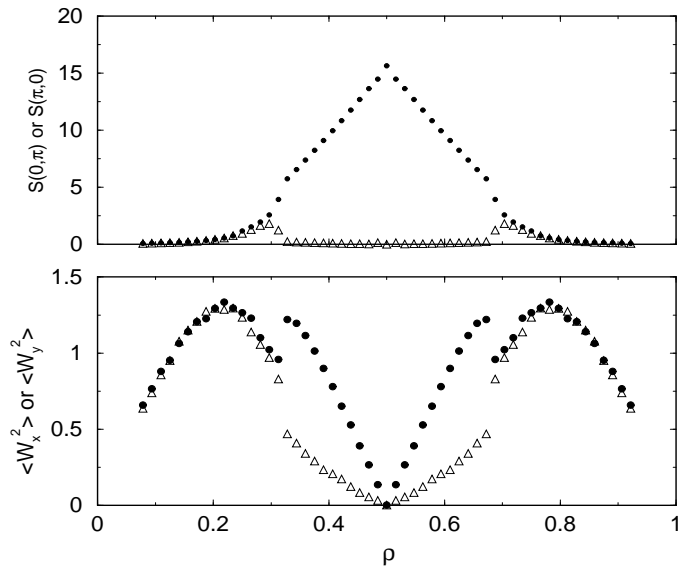,height=3.0in,width=3.5in,angle=-90}
\vskip-00mm
\caption{
Top: $S(0,\pi)$ (circles) and $S(\pi,0)$ (triangles). Bottom: $\langle
W^2_x\rangle$ (circles) and $\langle W_y^2\rangle$ (triangles). The
larger $\langle W^2\rangle$ (circles) is parallel to the stripes, the
lower is transverse. The system is $8\times 8$, $V_1=0$, $V_2=5$,
$\beta=6$ }
\end{figure}

Again we check the thermodynamic stability of the supersolid phase by
calculating $\rho$ as a function of $\mu$. This is shown in
Fig.~13. We see that the compressibility (slope) never becomes
negative indicating that phase separation is absent. In addition, we
see that the compressibility increases sharply at $\rho=0.3$ which is
the density at which stripes form (see Fig.~12). This indicates that,
contrary to the checkerboard case, the striped supersolid phase is
indeed thermodynamically stable and has a higher compressibility than
the superfluid phase.

It is worth emphasizing that the striped supersolid is {\it not}
merely a one dimensional superfluid phase along the channels created
by the stripes. If this were the case, the superfluid density
transverse to the stripes would be vanishingly small, which it is
not. In addition, it was shown in Ref.~\onlinecite{ggb2} that $\langle
a^{\dagger}({\vec r}) a({\vec r^{\,\, \prime}}) \rangle $ is finite as
$|{\vec r} -{\vec r^{\, \prime}}| \to \infty$ transverse to the stripes
in the supersolid phase.

\begin{figure}
\psfig{file=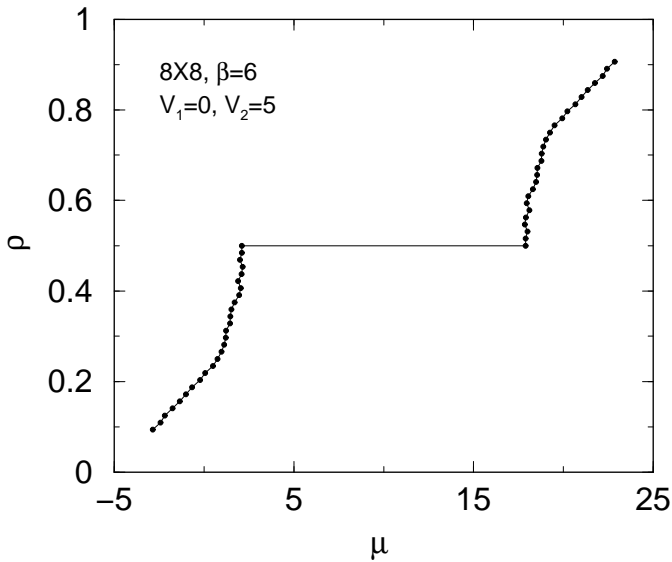,height=3.0in,width=3.5in,angle=-90}
\vskip-00mm
\caption{
Particle density, $\rho$, versus chemical potential, $\mu$. There is a
sharp increase in the compressibility as $\mu$ is increased when the
system goes from the superfluid to the supersolid phase.  }
\end{figure}

\begin{figure}
\psfig{file=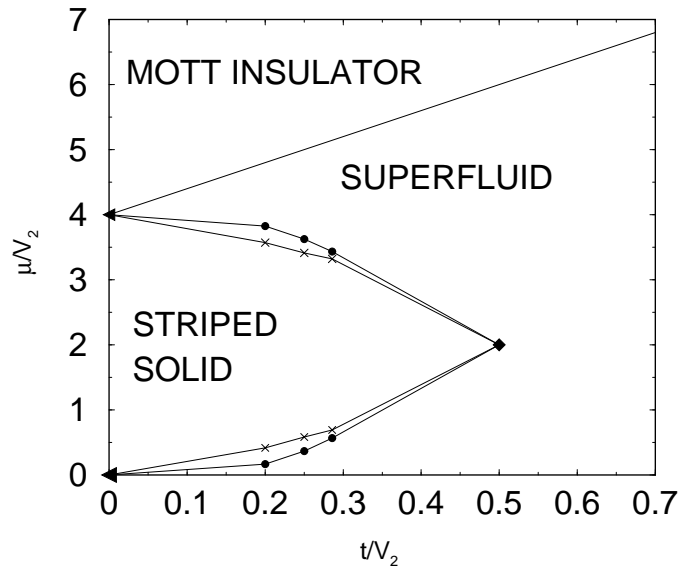,height=3.0in,width=3.5in,angle=-90}
\vskip-00mm
\caption{
The phase diagram for $V_1=0$. The narrow regions sandwiched
between SF and $(\pi,0)$ solid phases are the stable supersolid
phases.
}
\end{figure}

By repeating the simulations that led to Figs.~12 and 13, we map out
the phase diagram in the ($t/V_2,\mu/V_2$) plane. This is shown in
Fig.~14. The narrow regions sandwiched between SF and $(\pi,0)$ solid
phases are the stable supersolid phases.

As can be seen in Fig.~12, the transition from the SF to the
supersolid phase appears to be first order. The transition from the
supersolid to the striped solid phase is continuous as is seen from
the behavior of the superfluid density as half filling is approached.
We see from the lower part of Fig.~12 that both branches of the
superfluid density, parallel and transverse to the stripes, go to zero
smoothly as half filling is approached. In fact, both branches behave
like $|\rho-1/2|$, indicating a second order transition with a unit
exponent.

\section{Conclusions}

The boson--Hubbard model exhibits many fascinating quantum phases and
phase transitions.  In this paper, we have shown that the detailed
critical behavior at those transitions, that is, both the order of the
transitions and the critical exponents, can now be determined with
recently developed Quantum Monte Carlo
algorithms.  Our principal conclusion is that,
although a first order transition cannot be categorically ruled out at
the moment, the superfluid to checkerboard solid transition at
half--filling appears to proceed via a dynamical restoration of the
explicitly broken $O(3)$ symmetry, and is therefore in the same class
as the Heisenberg point. This confirms that, for the bosonic Hubbard
model at half filling, there is no supersolid phase between the
checkerboard solid and superfluid phases\cite{ggb3}, unlike what is
observed in other models\cite{vanott,roddick}. The details of the
finite temperature phase diagram are currently being worked
out\cite{finiteT}.

In addition, by examining the density histograms from a grand
canonical algorithm, we verified again that what was previously
thought to be a checkerboard supersolid phase, is in fact a phase
separated mixture of superfluid and solid regions.

As for the striped phases, we showed, from the hysteresis of the
energy and structure factor, that the superfluid to striped solid
transition, {\it at half filling}, is first order. The transition from
the superfluid phase to the striped supersolid phase (away from half
filling) also appears to be first order, in disagreement with
Refs.\onlinecite{frey-balents,pich-frey}. We also showed, from the
behavior of the superfluid density, that the transition from the
striped supersolid phase to the striped solid phase is second order
with the superfluid density vanishing as $\rho_s \sim |\rho - 1/2|$.

\vskip0.5in
\centerline{Acknowledgments}

We acknowledge useful conversations with E.W.~Fire.  This work was
supported by NSF--DMR--9985978. We also thank ETH-Zurich and HLRS
(Stuttgart) for very generous grants of computer time.


\begin{references}

\bibitem{FisherM} M.P.A. Fisher, P.B. Weichman, G. Grinstein, and
D.S. Fisher, Phys. Rev. {\bf B40}, 546 (1989).

\bibitem{cha}
M. Cha, M.P.A. Fisher, S.M. Girvin, M. Wallin, and
A.P. Young, Phys. Rev. {\bf B44}, 6883 (1991).

\bibitem{sorensen}
E.S. Sorensen, M. Wallin, S.M. Girvin, and A.P. Young,
Phys. Rev. Lett. {\bf 69}, 828 (1992).

\bibitem{ourprb5}
G.G.~Batrouni, B.~Larson, R.T.~Scalettar, J.~Tobochnik, and J.~Wang,
Phys.~Rev.~{\bf B48}, 9628 (1993).

\bibitem{runge}
K.J. Runge, Phys. Rev. {\bf B45}, 13136 (1992).

\bibitem{elesin}
V.F. Elesin, V.A. Kashurnikov, and L.A. Openov,
JETP Lett. {\bf 60}, 177 (1994).

\bibitem{kampfzim}
A.P. Kampf and G.T. Zimanyi,
Phys. Rev. {\bf B47}, 279 (1993).

\bibitem{ourprl2}
R.T.~Scalettar, G.G.~Batrouni, and G.T.~Zimanyi,
Phys.~Rev.~Lett.~{\bf 66}, 3144 (1991).

\bibitem{ourprb4}
G.T.~Zimanyi, P.A.~Crowell, R.T.~Scalettar,
and G.G.~Batrouni, Phys.~Rev.~{\bf B50}, 6515 (1994).

\bibitem{kanwal}
K.G.~Singh and D.S.~Rokhsar,
Phys. Rev. {\bf B46}, 3002 (1992).

\bibitem{spinflop} M. Kohno and M. Takahashi, Phys. Rev. {\bf B56},
                 3212 (1997).

\bibitem{ggb3}  G.G. Batrouni, R.T. Scalettar, G.T. Zimanyi
                 and A.P. Kampf, Phys. Rev.  Lett. {\bf 74} (1995)
                 2527; R.T. Scalettar, G.G. Batrouni, A.P. Kampf and
                 G.T. Zimanyi, Phys. Rev. {\bf B51} (1995) 8467.

\bibitem{ggb4} G.G. Batrouni and R.T. Scalettar,
Phys. Rev. Lett. {\bf 84} 1599 (2000).

\bibitem{monien} T.D.~Kuhner, S.R.~White, and H.~Monien,
Phys.~Rev.~{\bf B61}, 12474 (2000).

\bibitem{cha2}  M.C. Cha and S.M. Girvin,
Phys.~Rev.~{\bf B49}, 9794 (1994).

\bibitem{ourprl1} G.G.~Batrouni, R.T.~Scalettar, and G.T.~Zimanyi,
Phys.~Rev.~Lett. {\bf 65}, 176 (1990).

\bibitem{ourprb1+2+3} P.~Niyaz, R.T.~Scalettar, C.Y.~Fong, and
G.G.~Batrouni, Phys.~Rev.~{\bf B44}, 7143 (1991); G.G.~Batrouni and
R.T.~Scalettar, Phys.~Rev.~{\bf B46}, 9051 (1992); and P.~Niyaz,
R.T.~Scalettar, C.Y.~Fong, and G.G.~Batrouni, Phys.~Rev.~{\bf B50},
362 (1994).

\bibitem{Trivedi1}  W. Krauth and
N. Trivedi,  Europhys. Lett. {\bf 14}, 627 (1991).

\bibitem{ggb1} G.G. Batrouni and H. Mabilat, Comp.
                Phys. Comm. {\bf 121-122} 468 (1999).

\bibitem{ggb2} F. H\'ebert, G.G. Batrouni, and H. Mabilat,
                Phys. Rev. {\bf B61} 10725 (2000).

\bibitem{ggbdual} G.G. Batrouni, Nucl. Phys. {\bf B208} 467 (1982);
                 G.G. Batrouni and M. B. Halpern, Phys. Rev. {\bf D30}
                 1775 (1984).

\bibitem{sandvik} A. Sandvik, Phys. Rev. {\bf B59}, R14157 (1999).

\bibitem{troyer} A. Dorneich and M. Troyer, unpublished.

\bibitem{loop}  H.G. Evertz, G. Lana, and M. Marcu, Phys. Rev. Lett.
                 {\bf 70}, 875 (1993); U.-J. Wiese and H.-P. Ying,
                 Phys.\ Lett. {\bf A 168}, 143 (1992); Z. Phys.\ {\bf B
                 93} (1994) 147; N. Kawashima, J.E. Gubernatis, and
                 H.G. Evertz, Phys.\ Rev.\ {\bf B 50} 136 (1994);
                 N. Kawashima and J.E. Gubernatis, Phys.\ Rev.\ Lett.\
                 {\bf 73}, 1295 (1994), B. B. Beard and U.-J. Wiese,
                 Phys. Rev. Lett. {\bf 77}, 5130 (1996).

\bibitem{dishubsims}
Some additional studies of the boson--Hubbard model with disorder 
include:
N. Hatano, Int. J. Mod. Phys. {\bf C7}, 449 (1996);
N. Hatano, J. Phys. Soc. Japan {\bf 64}, 1529 (1995);
J. Kisker and H. Rieger,
Phys. Rev. {\bf B55}, 11981 (1997);
J. Kisker and H. Rieger,
Physica {\bf A246}, 348 (1997);

\bibitem{CP} E.L. Pollock and D.M. Ceperley,
              Phys. Rev. {\bf B30}, 2455 (1984) and Phys. Rev. {\bf
              B36}, 8343 (1987).  D.M. Ceperley and E.L. Pollock,
              Phys. Rev. Lett. {\bf 56}, 351 (1986).

\bibitem{vanott} A. van Otterlo, K.--H. Wagenblast, Phys. Rev. Lett.
                 {\bf 72}, 3598 (1994), A. van Otterlo {\it et al.},
                 Phys. Rev. {\bf B52}, 16176 (1995)

\bibitem{HIRSCH} J.E. Hirsch,
                Phys. Rev. Lett. {\bf 53} 2327 (1984).
                 Phys. Rev. {\bf B31} (1991) 6022.

\bibitem{CANNON}  J.W. Cannon, R.T. Scalettar, and E. Fradkin,
                 Phys. Rev. {\bf B44} (1991) 5995.

\bibitem{BOW}  P. Sengupta, A. Sandvik, and D.K. Campbell,
                cond--mat//0102141.

\bibitem{frey-balents} E. Frey and L. Balents, Phys. Rev. {\bf B55},
                        1050 (1997).

\bibitem{pich-frey} C. Pich and E. Frey, Phys. Rev. {\bf B57}, 13712
(1998).

\bibitem{finiteT} G. Schmid, M. Troyer and A. Dorneich, unpublished.

\bibitem{roddick} E. Roddick and D. Stroud, Phys. Rev. {\bf B51}, 8672
(1995).

\end{references}
\end{document}